\documentclass[11pt]{article}

\usepackage{xcolor}
\usepackage{amsfonts}
\usepackage{amsmath}
\usepackage{comment}
\usepackage{bbm}
\usepackage{multirow}
\usepackage{graphicx}
\usepackage{natbib}
\usepackage[margin = 1in]{geometry}
\newtheorem{theorem}{Theorem}

\title{Dynamic treatment regime characterization via a value function surrogate with an application to partial compliance}

\author{Nikki L. B. Freeman$^{1,*}$, 
Sydney E. Browder$^{2}$, 
Katharine L. McGinigle$^{3}$, \\ and
Michael R. Kosorok$^{1}$ \\
$^{1}$Department of Biostatistics, University of North Carolina, Chapel Hill, NC \\
$^{2}$Department of Epidemiology, University of North Carolina, Chapel Hill, NC \\
$^{3}$Department of Surgery, University of North Carolina, Chapel Hill, NC }

\begin{document}

\maketitle


\begin{abstract}
Precision medicine is a promising framework for generating evidence to improve health and health care. Yet, a gap persists between the ever-growing number of statistical precision medicine strategies for evidence generation and implementation in real world clinical settings, and the strategies for closing this gap will likely be context dependent. In this paper, we consider the specific context of partial compliance to wound management among patients with peripheral artery disease. Through the use of a Gaussian process surrogate for the value function, we expand beyond the common precision medicine task of learning an optimal dynamic treatment regime to characterization of classes of dynamic treatment regimes and how those findings can be translated into clinical contexts. 
\end{abstract}

%



%

\section{Introduction}
\label{s:intro}

In precision medicine, dynamic treatment regimes (DTRs), sequences of treatment rules over time that recommend treatments to patients based on observable, possibly time-evolving characteristics, are often the object of primary statistical interest. Common precision medicine analytic goals include inference for the value of a fixed regime, inference for comparisons between two or more regimes, and estimation of optimal regimes. Given the promise of optimal DTRs to improve treatment decisions and patient outcomes when employed, the focus on DTRs is well warranted. In clinical practice, however, DTRs and applications of precision medicine analytic treatment recommendation algorithms have not been widely adopted. 

Bridging the gap between precision medicine evidence and clinical decision support is important, context dependent, and may require expanding the focus of precision medicine analyses. For example, in the treatment of patients with lower limb wounds related to peripheral artery disease (PAD), a chronic vascular condition characterized by arterial blockages and poor limb perfusion, vascular surgeons attempt to save patients’ limbs by recommending a combination of wound care and surgery to improve blood flow. Often outpatient wound management in which patients complete regular wound clinic appointments over weeks to months is sufficient to heal the wound, and surgery to revascularize the limb can be avoided. With the goal of healing wounds and preserving limbs, determining which patients would benefit most from surgery and which patients can achieve equal outcomes with wound management alone is of great clinical interest. In the parlance of precision medicine, learning an optimal DTR for this single stage decision is of great importance.

However, given that the optimal treatment recommendation may depend on adherence to a rigorous outpatient schedule for wound management, embedded in this question is the need to understand if compliance matters in the revascularization decision-making calculus. If so, it is also important to know how much compliance is needed to heal wounds without surgery and how it relates to equivalent or near-equivalent clinical outcomes. These questions fall outside the objectives that traditionally have been pursued in precision medicine, yet they represent natural questions related to understanding and actually using optimal DTRs in a real world context. 

Strategies for dealing some of these issues in treatment effect estimation, such as imperfect compliance to treatment assignment, have been proposed including the use of instrumental variables (IVs) \citep{angrist_1996} and, more generally, principal stratification \citep{frangakis_2022}. In precision medicine, \citep{cui2021} offered an application of IVs to identify the value of a particular regime and to learn and optimal DTR under binary noncompliance. \cite{artman2020pc} took a principal stratification approach to estimate the value of regimes within strata in the setting of a Sequential Multiple Assignment Randomization Trial (SMART) \citep{lavori2000, lavori2004, murphy2009}. Still, some of the questions raised in the PAD setting have not been the focus of precision medicine yet are important to the translation of precision medicine into the clinical setting. 

In this paper, we bring a new perspective to the optimal DTR learning problem by focusing on tools for understanding DTRs and how to make them useful in practice which is particularly useful when considering real world complexities like partial compliance. Through the use of a Gaussian process surrogate for the value function, we (1) demonstrate the feasibility of employing Bayesian optimization for optimal policy search over a policy class with finite parameters, (2) exploit surrogate response surface methodology to characterize the approximate value of all policies in a finite-parameter policy space, and (3) describe how these findings can be translated to clinical contexts. The paper proceeds as follows: In Section \ref{section:background_notation} we set notation and formalize the setting. In Section \ref{s:approach}, we describe the general strategy for our approach. In Section \ref{s:simulation} we present simulation results for using Bayesian optimization for finding optimal DTRs, and in Section \ref{s:data_analysis} we present an application of our approach to the non-healing wounds in PAD setting that motivates our work. We conclude in Section \ref{s:discuss} with a discussion of our strategy, limitations, and future work. 

\section{Notation and setting}
\label{section:background_notation}
\subsection{Notation for the observed data}
We consider three time points $k = 0, 1, 2$ and two possible treatment options that are feasible to all patients $\mathcal{A} = \{0, 1\}$. We let $A_0$ and $A_1$ denote the treatment assignment at $k = 0$ and $k = 1$, respectively, and we further assume that all patients are initially assigned to treatment $0$, i.e. $A_0 \equiv 0$ for all patients. In the non-healing PAD wound setting, this is equivalent to giving patients an initial trial of non-surgical wound management instead of immediately opting for surgical intervention. Let $H_0 \equiv X_0 \in R^p$ denote the baseline covariates observed at the time of treatment initiation, $C_0 \in [0,1]$ denote compliance with the initial treatment assignment, and $X_1 \in R^q$ denote the updated covariates that arise in the interim after the initial treatment is assigned. At $k = 1$, the accumulated patient history is denoted by $H_1 = (X_0, A_0, C_0, X_1)$, where $\mathcal{H}_1$ denotes the set of all such histories, and is available for the treatment assignment decision between at $k = 1$. We let $C_1 \in [0, 1]$ denote the measured compliance with $A_1$, and let $Y$ denote the outcome of interest that is measured at $k = 2$. We assume our data contains i.i.d. sequences $(X_{0, i}, A_{0, i}, C_{0, i}, X_{1, i}, A_{1, i}, C_{1, i}, Y_i)$ for patients $i = 1, \ldots, n$.

\subsection{Potential outcomes and the precision medicine framework}

Our first goal is to learn an optimal dynamic treatment regime for the single decision of whether to continue with treatment $0$ or to switch to the treatment $1$ at the first decision point where treatment $0$ may be a lower intensity treatment with the potential for partial patient compliance and treatment $1$ is a higher intensity treatment with perfect compliance, i.e. one-sided partial compliance. In the setting we have thus far described, a dynamic treatment regime (DTR) $d$ is a map from the accumulated observed patient history at $k = 1$ to a treatment option, $d: \mathcal{H}_1 \longrightarrow \mathcal{A}_1$. Our interest is in regimes that are indexed by a finite parameter $\boldsymbol{\theta}$ which we denote by $\mathcal{D} = \{d(\boldsymbol{\theta}): \boldsymbol{\theta} \in \mathbf{\Theta} \subset \mathbb{R}^m, m < \infty\}$. Using the potential outcomes framework \citep{splawaNewyman1990, rubin1974}, let $C_1(a)$ and $Y(a, C_1(a))$ denote the compliance level and outcome we would observe, respectively, after assigning treatment level $a$ at $k = 1$. Similarly, let $C_1(d)$ and $Y(d, C_1(d))$ denote the compliance level and outcome, respectively, we would observe  after assignment treatment according to $d$ at $k = 1$. We define the value of a DTR $d$ as 
\[
    \mathcal{V}(d) = \mathbb{E}[Y(d, C_1(d))]
\]
where the expectation is with respect to the trajectory generated under policy $d$. An optimal dynamic treatment regime $d^{opt}$ is a regime such that $\mathcal{V}(d^{opt}) \ge \mathcal{V}(d)$  for all $d \in \mathcal{D}$. 

\subsection{Identification under one-sided partial compliance}
To identify the value of a DTR, which is defined in terms of potential outcomes, from the observed data, we must take care given that compliance to the assigned treatment is an intermediate variable on the causal pathway between treatment assignment and the outcome. In causal inference, one approach to handling post-treatment intermediate outcome variables is principal stratification, an approach that classifies individuals based on sets of the values of their potential intermediate variables \citep{frangakis_2022}. The canonical example in causal effect estimation is the case of two-sided noncompliance with binary treatments where strata are defined by the two potential compliance values for each individual. Letting $0$ and $1$ represent the two treatments and $C(0)$ and $C(1)$ denote the potential compliance levels for assignment to $0$ and $1$, respectively, the principal strata are $\mathcal{G}_{NC} = \{(0, 0), (0, 1), (1, 0), (1, 1)\}$. Sometimes these principal strata are referred to as ``never takers" $(0, 0)$, ``compliers" $(0, 1)$, ``defiers" $(1, 0)$, and ``always takers" $(1, 1)$. A benefit of principal strata is that they are unaffected by treatment assignment and thus can be viewed as a pretreatment variable, however principal strata present their own difficulties since they are unobserved. Treatment effects are often estimated conditional on a principal strata and are referred to as principal causal effects, e.g., $\mathbb{E}[Y(1) - Y(0) \vert G]$; identification of principal causal effects relies on structural or modeling assumptions, e.g. monotonicity or the exclusion restriction among others, with the plausibility of such assumptions depending on the specific problem context \citep{frangakis_2022, ding2017, imbens2015}. 

In our setting, the treatment is binary and we assume one-sided partial compliance (OSPC), that is $C_1(1) = 1$ and $C_1(0) \in [0, 1]$. Principal strata in this setting are thus $\mathcal{G} = \{(C_1(0), 1): C_1(0) \in [0, 1]\}$ which we can further simply as $\{C_1(0): C_1(0) \in [0, 1]\}$. This differs from the canonical noncompliance setting with binary treatment in the structural assumption that compliance is one-sided and that compliance may be partial rather than all-or-nothing. Further, in causal effect estimation, principal causal effects are often the target of estimation and provide an estimate of the treatment effect, for example, among those who comply. In contrast, to learn rules for optimal treatment recommendations, we are interested in rules that work across the target population, not just a targeted subgroup such as the complier subgroup. For that reason, using the principal stratification framework, we write the value of a DTR as 
\begin{align}\label{eq:dtr_value}
    \mathcal{V}(d) = \int_{0}^{1}\mathbb{E}[Y(d) \vert C_1(0) = c_1] dP_{C_{1}(0)}(c_1).
\end{align}
To identify (\ref{eq:dtr_value}) from the observed data, we assume that the Stable Unit Treatment Value Assumption (SUTVA) \citep{rubin1980} holds. We further assume that conditional on $H_1$, treatment assignment is independent of the potential outcomes $\{Y(0), Y(1), C_1(0)\}$ which we refer to as the no unmeasured confounders (NUC) assumption, and we assume positivity for each level of treatment $P(A_1 = a \vert H_1 = h_1) >0 \textrm{ for all } a \in \mathcal{A}, h_1 \in \mathcal{H}_1$. We further assume that $Y(1) \perp C_1(0) \vert H_1$ which is often referred to as principal ignorability (PI) \citep{jo2009}. In causal effect estimation, the PI assumption has motivated the use of principal scores for estimation of principal causal effects.

\begin{theorem}
For the binary treatment setting with OSPC, if the SUTVA, NUC, positivity and PI assumptions hold and if the class $\mathcal{P}_{\mathcal{H}_1} = \{P(C_1(0) \vert H_1 = h_1): h_1 \in \mathcal{H}_1\}$ is complete, then (\ref{eq:dtr_value}) is identifiable.
\end{theorem}

We note that in cases when the PI assumption is not plausible, it may be possible to identify (\ref{eq:dtr_value}) using an auxillary variable $W$ such that $Y(a) \perp W \vert C_1(0), H_1$ for $a = 0, 1$. We refer to this assumption as auxillary independence (AI) \citep{jiang2021}. In our setting, a candidate auxillary variable might be compliance to initial treatment to wound management $C_0$.

\section{Approach}
\label{s:approach}
In the setting described and given a parametric policy class $\mathcal{D}$ with parameters $\boldsymbol{\theta}$, our objective is to optimize $\mathcal{V}$ over $\boldsymbol{\theta}$ and to characterize the estimated values of the policies in $\mathcal{D}$. In this section we propose a simple strategy for estimating the value of a fixed DTR that incorporates partial compliance information. We follow by proposing the use of a Gaussian process surrogate model for the value function to aid in both the optimization goal and the characterization goal. 

\subsection{Value estimation with partial compliance}
We use a simple imputation-based estimator for (\ref{eq:dtr_value}). $C_1(0)$ is missing for those assigned to $A=1$. To impute this missing value, we propose estimating $C_1(0) \sim H_1$ from the class of complete distributions among those with $A_1 = 0$. Under the assumption of principal ignorability, we can then impute $C_1(0)$ from the posterior predictive distribution of $C_1(0) \vert H_1$. Now that $C_1(0)$ is complete, we can estimate a model $Y \vert C_1(0), H_1, A$. To estimate the value of a fixed regime, we take an approach similar to (\cite{chakrabortyBook}, Chapter 6.2):
For a fixed, finite-parameter regime $d(\boldsymbol{\theta})$
\begin{enumerate}
    \item Draw a large number of observations from the sample with replacement.
    \item For each observation, if $C_1(0)$ is unobserved, draw $C_1(0)$ from the posterior predictive distribution $C_1(0)\vert H_1$.
    \item Fix the treatment vector such that $A$ is consistent with the fixed regime of interest $d$.
    \item For each observation, draw from the posterior predictive distribution of $Y \vert C_1(0), H_1, A$.
\end{enumerate}
To get a draws from the distribution of mean counterfactual outcomes under regime $d$, we (5) average the draws in step $(4)$ and repeat steps (2)-(5) a large number of times. One then may use the mean or median of the mean counterfactual outcomes under regime $d$ as an estimate for (\ref{eq:dtr_value}).

\subsection{General approach for DTR class characterization}

Because one of our goals is to characterize policies in a policy class by their value, an important aspect of our approach is estimation of the value function for a fixed policy, i.e. policy evaluation. A number of strategies for policy evaluation have been proposed including inverse probability weighting \citep{robins2000}, augmented inverse probability weighting \cite{zhang_2012_robust}, and G-computation \citep{robins1986}. While policy evaluation is important for some precision medicine questions, it is not immediately practical for finding optimal policies among a large policy class nor for characterizing the value of an entire policy class. For example, one could achieve these goals using grid search over the policy class which is impractical except for the simplest of policies and settings. 

Our approach starts with noting that for a finite parameter policy class the value function can be viewed as a function mapping from the policy parameter space to a real number $\mathcal{V}: \boldsymbol{\theta} \longrightarrow \mathbb{R}$. While the value function is generally unknown, we conceptualize policy evaluation as a strategy for sampling the value function at a given policy parameter value. We then model $\mathcal{V}$ using a Gaussian process regression (GPR) model, that is, we place a Gaussian process (GP) prior on $\mathcal{V}$ which we update by sampling $\mathcal{V}$ to yield a posterior predictive distribution that can be used to approximate the value at unsampled policy parameters.

A Gaussian process $W$ indexed by a set $T$ is a stochastic process such that finite-dimensional distributions of $(W(t_1), \ldots, W(t_m))$, for all $t_1, \ldots, t_m \in T$, $m \in \mathbb{N}$ are jointly Gaussian distributed and completely specified by mean function $\mu:T \longrightarrow \mathbb{R}$ and covariance kernel $K: T \times T \longrightarrow \mathbb{R}$ with $\mu(t) = \mathbb{E}[W(t)]$ and $K(s, t) = cov(W(s), W(t))$ \cite{gpbook, ghosalbook}. As random functions, Gaussian processes can serve as priors on function spaces and from this view can be considered as a flexible method for nonparametric Bayesian inference directly on a function space and is often referred to as Gaussian process regression (GPR). By Mercer's theorem, GPR can be viewed as Bayesian linear regression with possibly infinitely many basis functions \cite{gpbook, ghosalbook}. 

We write the GPR model for $\mathcal{V}$ as
\begin{align*}
    \mathcal{V}(\boldsymbol{\theta}) \sim \mathcal{GP}(\boldsymbol{0}, K(\boldsymbol{\theta}, \boldsymbol{\theta}^{\prime}))    
\end{align*}
where $\mathcal{GP}$ denotes Gaussian process prior and we have taken the mean function to be zero. Assuming that we have a strategy for evaluating policies in $\mathcal{D}$, e.g. inverse-probability weighting, augmented inverse-probability weighting, G-computation, we can select a small number of policies $\boldsymbol{\theta}_1, \ldots, \boldsymbol{\theta}_n$, where the selection of points are by design, such as a space-filling design, to evaluate and yield corresponding estimates of the value $\widehat{\mathcal{V}}^{(1)}, \ldots, \widehat{\mathcal{V}}^{(n)}$. The $n$ evaluation pairs $\{(\boldsymbol{\theta}, \widehat{\mathcal{V}}^{(1)}), \ldots, (\boldsymbol{\theta}, \widehat{\mathcal{V}}^{(n)})\}$ can be used to update the prior and yield the predictive distribution for predicting the value for unevaluated policies, i.e. policies that correspond to $\boldsymbol{\theta} \notin \{\boldsymbol{\theta}_1, \ldots, \boldsymbol{\theta}_n\}$. In particular, if we let $\widehat{\mathcal{V}}^{(\cdot)}$ denote $(\widehat{\mathcal{V}}^{(1)}, \ldots, \widehat{\mathcal{V}}^{(n)})^\top$, $\bar{\boldsymbol{\theta}}$ denote $(\boldsymbol{\theta}_1, \ldots, \boldsymbol{\theta}_n)^\top$, $\boldsymbol{\theta}_{n+1}$ denote the parameter indexing a policy whose value we wish to approximate based on our model, and $\widehat{\mathcal{V}}^{(n+1)}_{\ast}$ as the value corresponding to $\boldsymbol{\theta}^{(n+1)}$, then we can write the joint distribution of the evaluation pairs and the value function at $\boldsymbol{\theta}^{(n+1)}$ as
\begin{align*}
    \begin{bmatrix}
        \widehat{\mathcal{V}}^{(\cdot)} \\
        \widehat{\mathcal{V}}^{(n+1)}_{\ast}
    \end{bmatrix}
    \sim
    \mathcal{N}
    \begin{pmatrix}
        \boldsymbol{0}, 
        \begin{bmatrix}
            K(\bar{\boldsymbol{\theta}}, \bar{\boldsymbol{\theta}}) + \widehat{\sigma}^2_n I & K(\bar{\boldsymbol{\theta}}, \boldsymbol{\theta}^{(n+1)}) \\
            K(\boldsymbol{\theta}^{(n+1)}, \bar{\boldsymbol{\theta}}) & K(\boldsymbol{\theta}^{(n+1)}, \boldsymbol{\theta}^{(n+1)})
        \end{bmatrix}
    \end{pmatrix}
\end{align*}
where $\widehat{\sigma}_{n}^2$ is the estimated variance for the value estimates. The conditional distribution of $\widehat{\mathcal{V}}^{(n+1)}_{\ast}$ given $\bar{\boldsymbol{\theta}}$, $\widehat{\mathcal{V}}^{(\cdot)}$, and $\boldsymbol{\theta}^{(n+1)}$ follows from basic distribution theory for the multivariate Gaussian distribution and is given by
\begin{align}\label{eq:gp_predictive}
    \widehat{\mathcal{V}}^{(n+1)}_{\ast} \vert \bar{\boldsymbol{\theta}}, \widehat{\mathcal{V}}^{(\cdot)}, \boldsymbol{\theta}^{(n+1)} &\sim \mathcal{N}(K(\boldsymbol{\theta}^{(n+1)}, \bar{\boldsymbol{\theta}})[K(\bar{\boldsymbol{\theta}}, \bar{\boldsymbol{\theta}}) + \widehat{\sigma}^2_n I]^{-1}\widehat{\mathcal{V}}^{(\cdot)}, \text{cov}(\widehat{\mathcal{V}}^{(n+1)}_{\ast})), \text{ where}\\ \nonumber
    \text{cov}(\widehat{\mathcal{V}}^{(n+1)}_{\ast}) &= K(\boldsymbol{\theta}^{(n+1)}, \boldsymbol{\theta}^{(n+1)}) - K(\boldsymbol{\theta}^{(n+1)}, \bar{\boldsymbol{\theta}})[K(\bar{\boldsymbol{\theta}}, \bar{\boldsymbol{\theta}}) + \widehat{\sigma}^2_n I]^{-1} K(\bar{\boldsymbol{\theta}}, \boldsymbol{\theta}^{(n+1)}).
\end{align}
We use the estimated GPR model in two ways, first to help us find the optimal DTR and second, to help us characterize the policy class at large. We describe these two functions for the estimated GPR model in the next two subsections.

\subsection{Optimal DTR learning}
Bayesian optimization is a machine learning approach to solving problems of the form $\underset{x \in \mathcal{X}}{\max} f(x)$ where $\mathcal{X} \in \mathbb{R}^d$ is the feasible set, typically a simple set like a hyper-rectangle or $d-$dimensional simplex, and $f$ is the objective function. Typically, Bayesian optimization methods employ a surrogate for the expensive-to-evaluate function $f$ and define an acquisition function from the surrogate to guide which points to evaluate next. Different surrogates and different acquisition functions correspond to different Bayesian optimization strategies, and the fundamental idea is that the acquisition function enables computationally efficient search for optima over $\mathcal{X}$ \citep{frazier2018}.

In the optimal DTR learning context, the objective function we want to optimize is the value function, the GPR model serves as our surrogate for the value function, and an acquisition function is used to guide which value of $\boldsymbol{\theta}$ to evaluate next. Multiple acquisition functions and approaches to Bayesian optimization have been proposed, each with their own rationale and varying theoretical guarantees. One popular approach is the Expectation Improvement (EI) algorithm \citep{schonlau1998} whose acquisition function can be viewed as balancing exploration and exploitation. EI may be implemented to learn an optimal DTR as follows:

\begin{enumerate}
    \item Select $b$ initial values of $\boldsymbol{\theta} \in \mathbf{\Theta}$ using a space-filling design; label them as $\boldsymbol{\theta}^{(1)}, \ldots, \boldsymbol{\theta}^{(b)}$ and the corresponding policies they index as $d^{(1)}, \ldots, d^{(b)}$. Evaluate the value function under each of the policies $d^{(1)}, \ldots, d^{(b)}$, and label them as $\mathcal{V}^{d^{(1)}}, \ldots, \mathcal{V}^{d^{(b)}}$.
    \item Fit a Gaussian process to the counterfactual draws of $-\mathcal{V}^{d^{(1)}}, \ldots, -\mathcal{V}^{d^{(b)}}$ over the policy parameter space $\boldsymbol{\Theta}$ 
    \begin{align}
        -\mathcal{V} \sim \mathcal{GP}(m(\boldsymbol{\theta}), k(\boldsymbol{\theta}, \boldsymbol{\theta})).
    \end{align}
    Here, we negate the draws from the value because EI is traditionally written as a minimization routine.
    \item Compute the expected improvement $EI$
    \begin{equation}
        EI(\boldsymbol{\theta}) = (f^{(b)}_{min} - \mu_{(b)}(\boldsymbol{\theta})) \Phi\left( \frac{f^{(n)}_{min} - \mu_b(\boldsymbol{\theta})}{\sigma_b(\boldsymbol{\theta})} \right) + \sigma_{b}(\boldsymbol{\theta}) \phi \left( \frac{f^{(b)}_{min} - \mu_{b}(\boldsymbol{\theta})}{\sigma_{b}(\boldsymbol{\theta})} \right)
    \end{equation}
    where $\Phi(\cdot)$ is the cumulative distribution function for the standard normal distribution, $\phi(\cdot)$ is the probability density function for the standard normal distribution, $\mu_{(b)}(\cdot)$ is the predictive mean function of the fitted Gaussian process, and $\sigma_{b}(\cdot)$ is the predictive variance function of the fitted Gaussian process.
    \item Find the value of $\boldsymbol{\theta}$ that maximizes EI and evaluate the value function for the associated policy.
    \item Let $b \leftarrow b+1$ and repeat Step 3 and Step 4 until maximum EI small or the sampling budget is met.
\end{enumerate}
The estimated optimal DTR is the policy indexed by $\boldsymbol{\theta}^{\ast} = \max \{\boldsymbol{\theta}^{(1)}, \ldots, \boldsymbol{\theta}^{(b)}\}$ so that $d^{opt} = d(H_1; \boldsymbol{\theta}^{\ast})$.


\subsubsection{Policy class characterization}
While optimization of the value function may yield one or more points in the policy parameter space that are optimal, solely optimizing the value function alone cannot tell us how quickly the value changes with a change in one or more of the parameters. Moreover, in the case of value functions where the optima occur in flat, non-peaky regions of the value function, it may be the case that an entire set of decision rules yield nearly equivalent estimated values and equivalent benefit in terms of clinical significance. For example, using the predictive mean function (\ref{eq:gp_predictive}) of the fitted Gaussian process, we can quickly approximate the value of policies in the entire policy class and thereby identify multiple optima and the points in policy parameter space where benefits begin to drop off. With the ability to sample from the posterior predictive distribution, a wide range of numerical and visual summaries can be computed which may further aid in characterizing DTRs. 

\section{Simulation}
\label{s:simulation}

We conducted simulation studies to examine the performance of our approach for DTR class characterization, both learning an optimal DTR and for approximating the value of DTRs in a given policy class which are the novel aspects of our approach. For Simulation Setting 1, we drew $U \sim Uniform(0, 1)$ and set $X = wU$ where $w$ is a positive scale parameter. We drew $A \sim Bern(0.5)$, and set $Y = \gamma_0 + \gamma_1 x + A[I(x_1 <0.75 \text{ or } x_1 > 0.25) + 0.5*x*(1 - I(x_1 <0.75 \text{ or } x_1 > 0.25))]$. For Simulation Setting 2, we drew $U$, $A$, and $X$ in the same way as Simulation Setting 1 and set $Y = \gamma_0 + \gamma_1 x_1 + A*cos(2 \pi x_1)$. Simulation Setting 3 is the same as Simulation Settings 1 and 2 except for setting $Y = \gamma_0 + \gamma_1 x + A*\cos(4x\pi)$. In each simulation setting, we searched over the policy class $\mathcal{D} = \{d(x; \beta): d(x; \beta) = \mathbbm{1}(x < \beta_1 \text{ or } x > \beta_2), \text{ for } \beta_1, \beta_2 \in [0, 1]\}$.

\begin{figure}
    \centering
    \includegraphics[width = \textwidth]{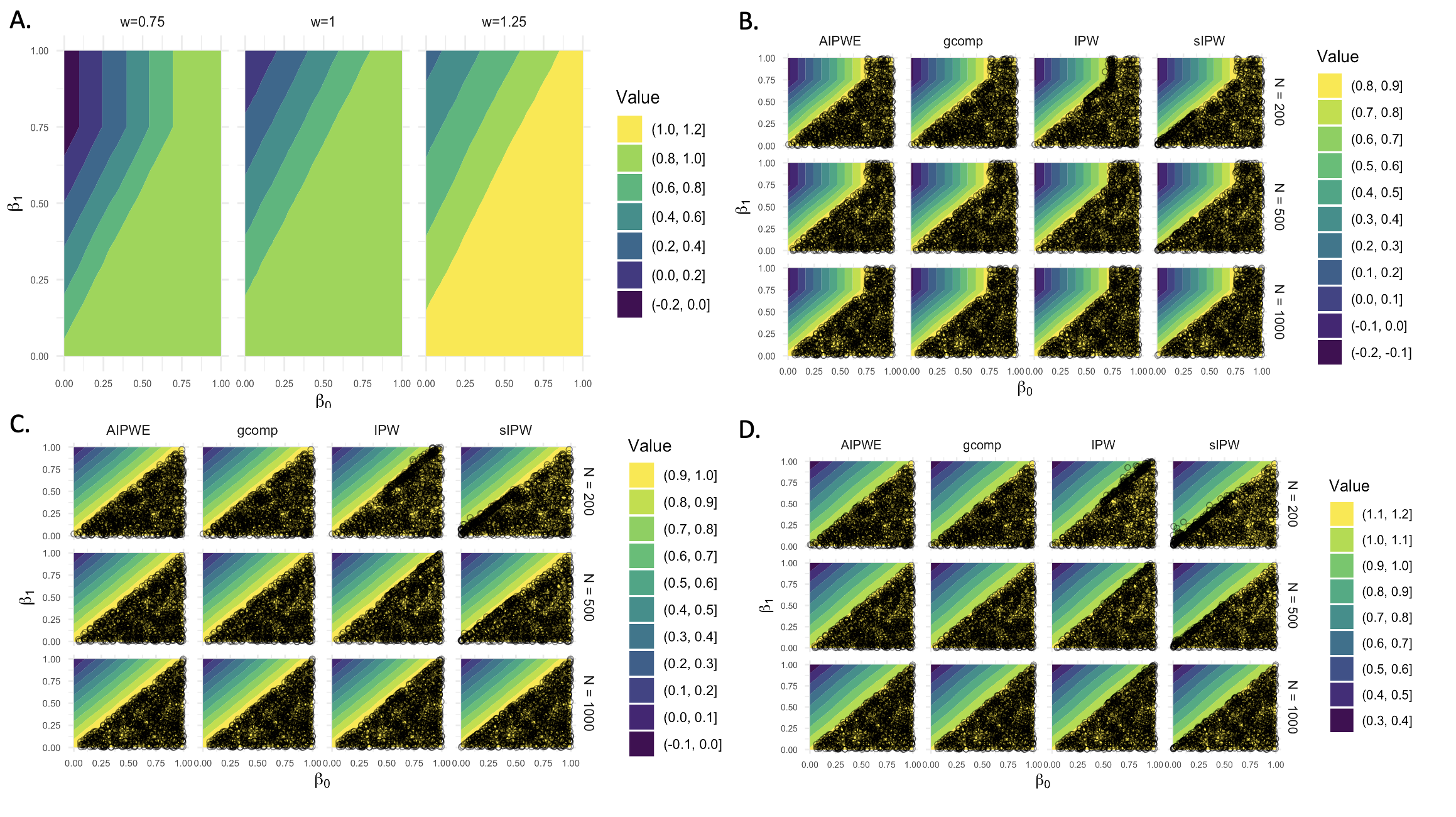}
    \caption{Simulation Setting 1}
    \label{fig:sim_setting1}
\end{figure}

\begin{figure}
    \centering
    \includegraphics[width = \textwidth]{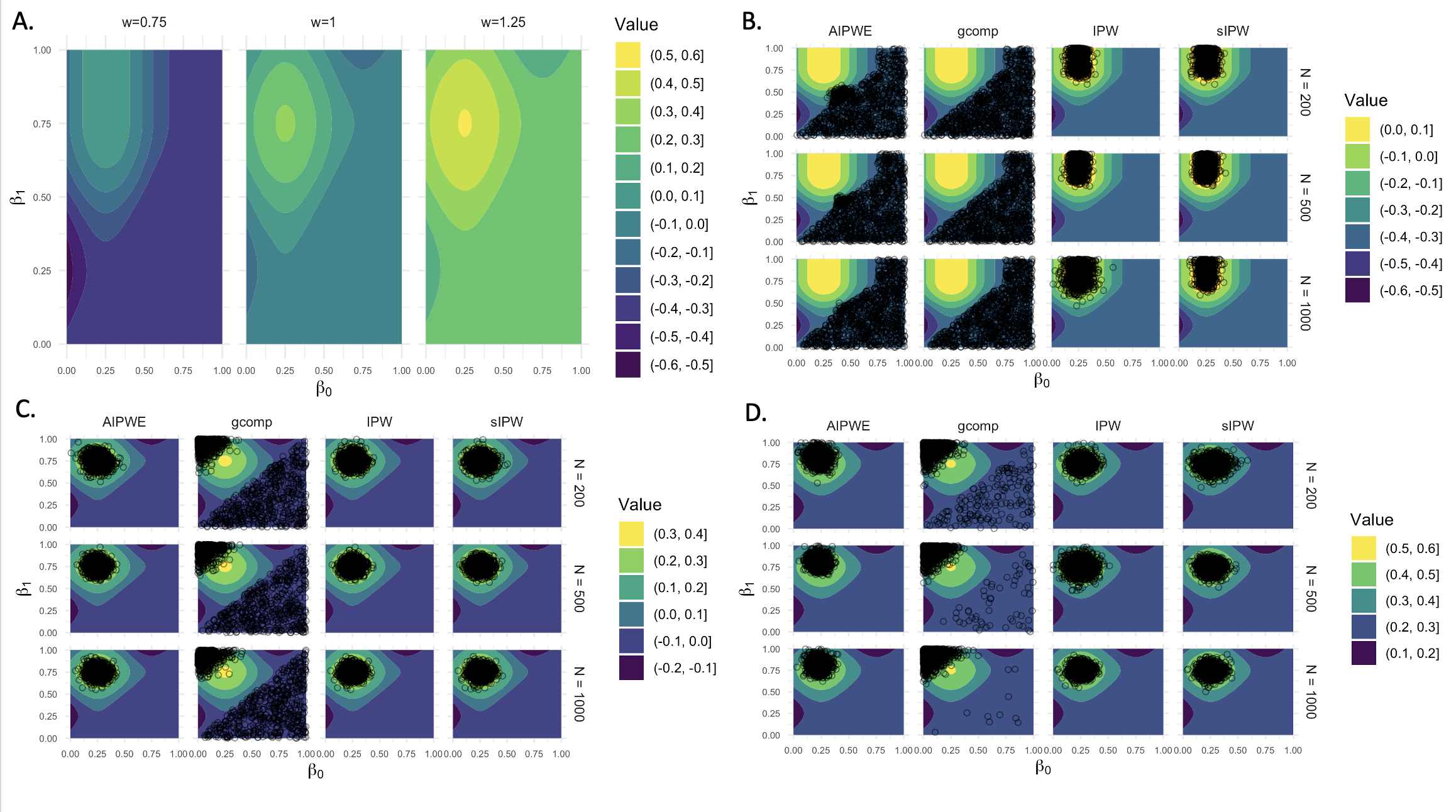}
    \caption{Simulation Setting 2}
    \label{fig:sim_setting2}
\end{figure}

\begin{figure}
    \centering
    \includegraphics[width = \textwidth]{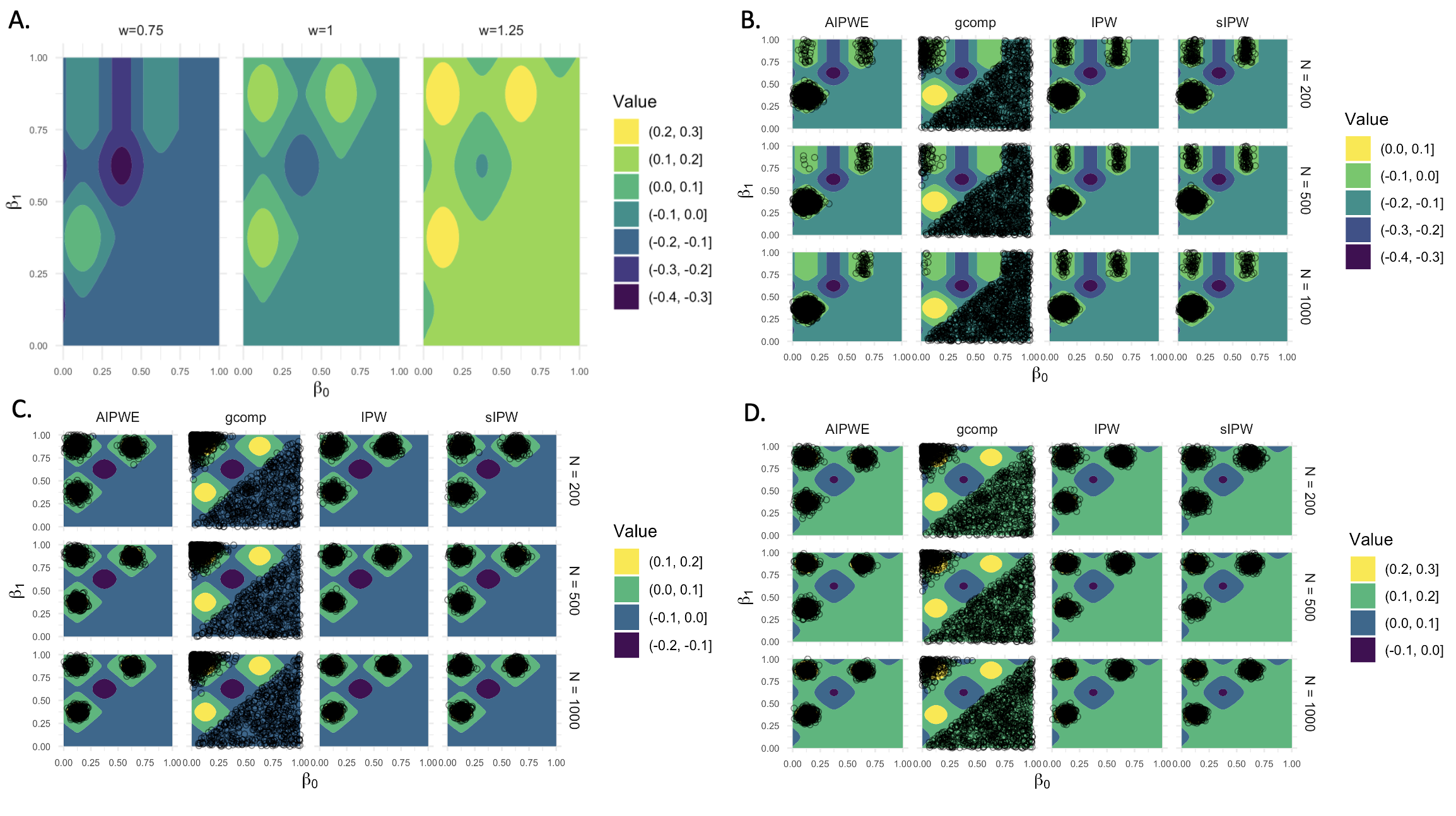}
    \caption{Simulation Setting 3}
    \label{fig:sim_setting3}
\end{figure}

For each simulation setting, we considered $w \in \{0.75, 1, 1.25\}$. The true value of policies in $\mathcal{D}$ for each setting and each value of $w$ is visualized by the contour plots of the value as a function of policy parameters $\beta_0$ and $\beta_1$ in Panel A of Figures \ref{fig:sim_setting1}, \ref{fig:sim_setting2}, and \ref{fig:sim_setting3}. In Simulation Setting 1, there are many optimal policies for each value of $w$. In Simulation Setting 2, the optimal policies are in a narrower range of $\beta_0$ and $\beta_1$ and the optimal policy region is more complex than that of Simulation Setting 1. Simulation Setting 3 is characterized by multiple peaks.

Each simulation study was characterized by the setting and policy class under examination, the sample size $N = 200, 500, 1000$, $w = 0.75, 1, 1.25$, and the policy evaluation method employed: IPW, stabilized IPW, G-computation (gcomp), or AIPWE. For the G-computation and AIPWE value estimators, we misspecified the conditional mean models; for IPW, sIPW, and AIPWE we used the true sampling weights. For each study we computed 1000 simulation runs. For each run we initially evaluated 50 random policies to fit a GPR model for the value function and then used EI to evaluate 50 more policies. The estimated optimal policy is the policy, among those evaluated, with the highest value. To characterize the policy class, we fixed a fine grid over the policy parameter space and used the posterior predictive mean function of the GPR model (Matern kernel + white noise) to approximate the value of policies on the grid. To assess the performance of the GPR and EI approach to learn an optimal policy, we computed the mean square error (MSE) between the estimated value and the oracle value over the 1000 simulation runs; full results are given in the appendix and visualized in Panels B, C and D of Figures \ref{fig:sim_setting1}, \ref{fig:sim_setting2}, and \ref{fig:sim_setting3}. To assess the performance of the GPR to approximate the policy class, we computed the $L^1$ norm and $L^2$ norm between the posterior predictive mean of the value function GP surrogate and the true value at each point on the aforementioned fine grid; full results are included in the appendix.

In Panels B, C, and D of Figures \ref{fig:sim_setting1}, \ref{fig:sim_setting2}, \ref{fig:sim_setting3} the background contours correspond to the true value of the policy indexed by $\beta_0$ and $\beta_1$ and each open point represents the estimated value of the optimal DTR from a single simulation run. By visual inspection, in Simulation Setting 1, each value estimation method yielded optimal policies in the region of the policy parameter space that corresponds to the highest value, meaning that optimal policies were found (MSE for each case $\le 0.007$). In Simulation Settings 2 and 3, the grossly misspecified conditional mean model for the G-computation value estimator yielded sub-optimal policies as well as in the case of AIPWE in Simulation Setting 2 for $w=0.75$. This is not surprising since our strategy hinges on having a good estimator for the value. IPW and stabilized IPW perform well across settings, with estimated optimal policies concentrating on the regions of the policy parameter space with the highest value. 

\begin{figure}
    \centering
    \includegraphics[width = \textwidth]{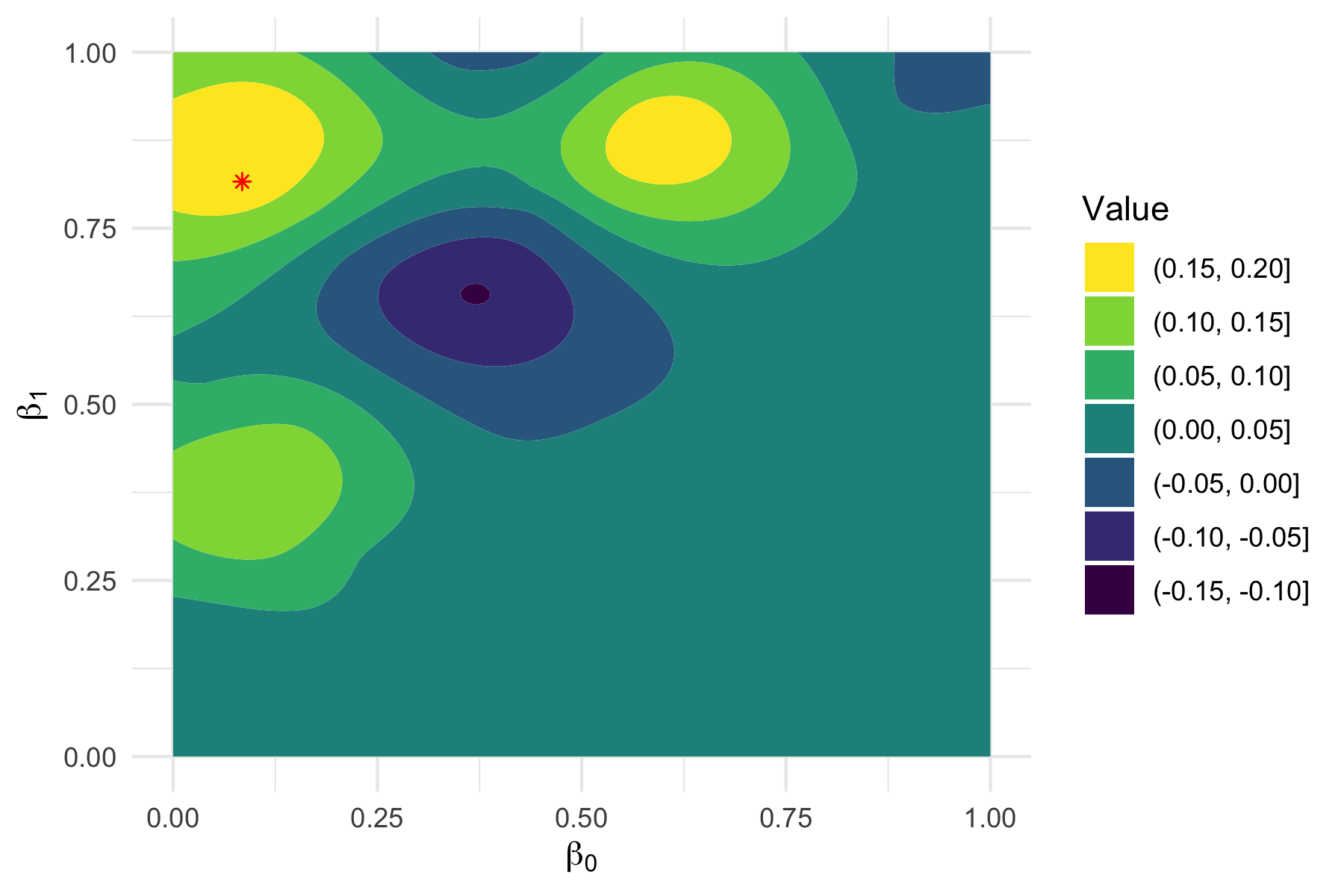}
    \caption{Example GPR surface from Simulation Setting 3 (sIPW with $N = 200$, $w = 1$)}
    \label{fig:exemplar}
\end{figure}

One particularly promising aspect of using a GPR surrogate for the value function is the ability to characterize the values for policies in an entire policy class, thereby finding cases where there are multiple optimal or near-optimal policies that would yield similarly clinically significant results. We demonstrate this with the first simulation run for Simulation Setting 3 with $w= 1$ and $N=200$, a setting with 3 distinct sub-classes of policies that yield the highest values. In Figure \ref{fig:exemplar}, we plot as contours the predicted mean values across the policy parameter space after training a GPR model (Matern kernel + white noise) with the evaluation points used for optimization. The red star represents the estimated optimal policy. In this case, the estimated GPR surrogate identifies the regions of the policy parameter space that yields policies with the highest values where the distance between the true value surface and surrogate differs by 0.04 in the $L^2$ sense.

\section{Dynamic treatment regimes for partially compliant ambulatory wound care patients}
\label{s:data_analysis}
\subsection{Setting and data}
PAD is narrowing of the arteries outside of the heart and brain. Because diseased arteries result in reduced blood flow to the lower limbs, many patients develop wounds that are slow to heal, painful, and at risk of becoming infected or necrotic. PAD patients are complex with multiple comorbidities such as diabetes, coronary artery disease, cerebrovascular disease, and hypertension, and limb loss rates (amputation) and mortality rates are high. For patients with PAD with non-healing wounds, treatment may consist of outpatient wound management, resvascularization surgery such as a bypass or an endovascular procedure, or a combination of both. Given the multimorbidity of this patient population, non-surgical treatment of non-healing wounds via wound management may be preferable to surgery when it is possible to heal wounds with wound care alone \citep{browder2022}, however, wound management requires patients to attend scheduled appointments to receive treatment until the wound is healed. Thus, it is useful to know (1) given a trial of wound management, which patients would benefit from more intensive treatment (surgery) and which patients would likely benefit from continued wound management, and (2) how compliance to a trial of wound management should drive the decision to switch to a more intensive strategy to heal patients' wounds. 

We analyzed a single-institution cohort of patients with PAD that were referred to outpatient wound care. We extracted patient characteristics, including age, sex, race, comorbidities, number of wounds, wound location, and initial wound size from the electronic health records of patients with PAD that were seen at UNC hospitals between 2013 and 2018. We linked the EHR data to administrative scheduling data from a wound management clinic in the UNC system. Patients were considered compliant with wound management if they completed their scheduled appointments and non-compliant if they cancelled or missed their scheduled appointments. We considered the first month following the first wound management appointment as a trial of wound management and compliance during that period as baseline compliance. For those with wounds that remained unhealed at 1 month after initiating wound management and with limbs that had not been revascularized during the first month of treatment, we sought to learn an optimal dynamic treatment of the form ``Revascularize along with continued wound care if baseline compliance is less than or equal to $\theta_1$ or if the initial wound size of the largest wound a patient has is greater than $\theta_2$" that optimizes the outcome of being healed, alive, and with an intact limb (not amputated) at 6 months after the initiation of wound management. If we let $C_0$ denote the proportion of completed wound care visits in the first month of wound management and $W_0$ denote initial wound size, then we can write the class of policies under examination as $\mathcal{D} = \{d(C_0, W_0; \theta_1, \theta_2) = \mathbbm{1}(C_0 < \theta_1 \text{ or } W_0 > \theta_1): \theta_1 \in [0, 1], \theta_2 \in (0, 100]\}$ where $d(C_0 = c_0, W_0 = w_0; \theta_1, \theta_2) = 1$ recommends revascularization and wound care for a particular baseline compliance $c_0$ and intial wound size $w_0$ and $d(C_0 = c_0, W_0 = w_0; \theta_1, \theta_2) = 0$ recommends continuing with wound care alone. We define compliance after the treatment decision as in the trial of wound care period, that is, compliance relates only to wound management and determined from the administrative wound clinic data. 

\subsection{Learning an optimal dynamic treatment regime}
Using the notation and under the assumptions in Section \ref{section:background_notation},  compliance strata $C_1(0)$ was modeled as
\begin{align*}
    C_1 \vert H_0, A=0 &\sim \textit{Truncated Normal}(\mu_{C_1}, \sigma_{C_1}, 0, 1; \boldsymbol{\beta}_{C_1}) \\
    \boldsymbol{\beta}_{C_1} \sim MVN(\boldsymbol{0}, 3I),& \quad
    \sigma_{C_1} \sim Half Cauchy(5)
\end{align*}
where $\mu_{C_1}$ is a linear predictor with parameters $\boldsymbol{\beta}_{C_1}$ and covariates are baseline patient and wound characteristics and compliance level during the first month of wound care, $I$ denotes the identity matrix, $\sigma_{C_1}$ is the standard deviation, and $0$ and $1$ are lower and upper boundaries, respectively, of the support of the truncated distribution. The binary outcome of being healed, alive, and with the limb intact at 6 months after initiating wound care, $Y$, was modeled as
\begin{align*}
    P(Y = 1) \sim Bernoulli(g^{-1}(\eta(H_1, \boldsymbol{\beta}_{Y}))), \quad \boldsymbol{\beta}_Y \sim MVN(\boldsymbol{0}, \Sigma_Y)
\end{align*}
where $g$ is the canonical link function, $\eta$ is the linear predictor with parameters $\boldsymbol{\beta}_Y$ and covariates $H_1$ that include baseline patient and wound features, baseline compliance, treatment, and compliance strata. $\Sigma_Y$ is a diagonal matrix and a hyperparameter for the prior for $\boldsymbol{\beta}_Y$. Both models were fit using Hamilton Monte Carlo and the No U-Turn Sampler in Python version 3.9.7 with modules PyMC3 version 3.11.4, ArviZ version 0.11.2, and Bambi version 0.7.1. Trace plots, sampler statistics, and model summary statistics were analyzed to assess chain convergence and model fit. 

To learn an optimal dynamic treatment regime, we used the expectation improvement algorithm with 25 initial random starts and a budget of 25 evaluations to minimize the negative of the value function over the policy parameter space $[0, 1] \times (0, 100]$. To evaluate the value function, we first sampled counterfactuals by constructing a hypothetical population of 10,000 samples with replacement from the observed data and drew $C_1 \vert H_1, C_0$ and $Y\vert H_2$ from their respective posterior predictive distributions where treatments were fixed as according to the rule implied by $\theta_1$ and $\theta_2$. Repeating this 30 times and averaging over the counterfactual outcomes yielded a draw from distribution of the value function under the given rule. The Bayesian optimization step was conducted in Python version 3.9.7 with the module scikit-optimize version 0.9.0.

\subsection{Results and characterization of dynamic treatment regimes}
\label{sec:real_data_resluts}
Our analysis yielded an optimal DTR: $\widehat{d}^{\text{opt}}(C_0, W_0; \widehat{\theta}_1, \widehat{\theta}_2) = \mathbbm{1}(C_0 < 0.23 \text{ or } W_0 > 89.8)$. This means that for PAD patients with wounds who after a 1-month trial of wound management have not healed their wound, surgical revascularization in addition to continued wound care should be recommended for those whose completed less than 1/4 of their wound management appointments during the first month of outpatient wound care and for those whose initial wound size is larger than 89.8 cm$^2$. In the observed data, 34.2\% of patients were healed, alive, and with their limb intact at 6 months after initializing wound care. The estimated value of $\widehat{d}^{\text{opt}}$ is 46.9\%, a 12.2\% improvement.

\begin{figure}
    \centering
    \includegraphics[width = 8cm]{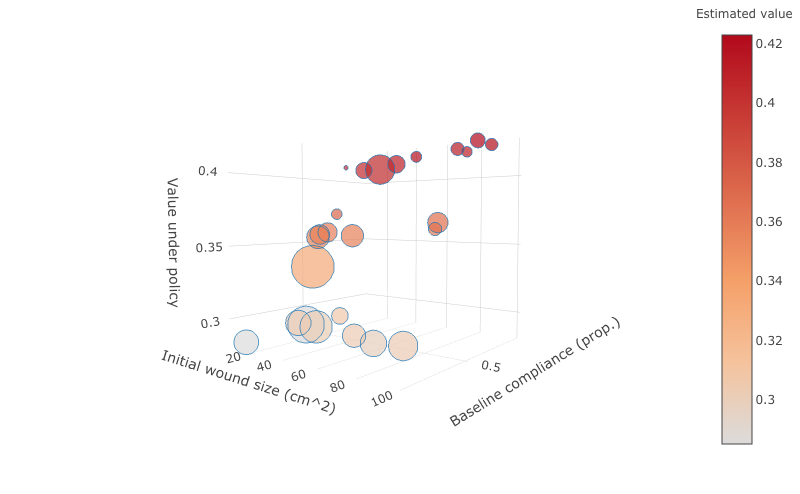}
    \includegraphics[width = 8cm]{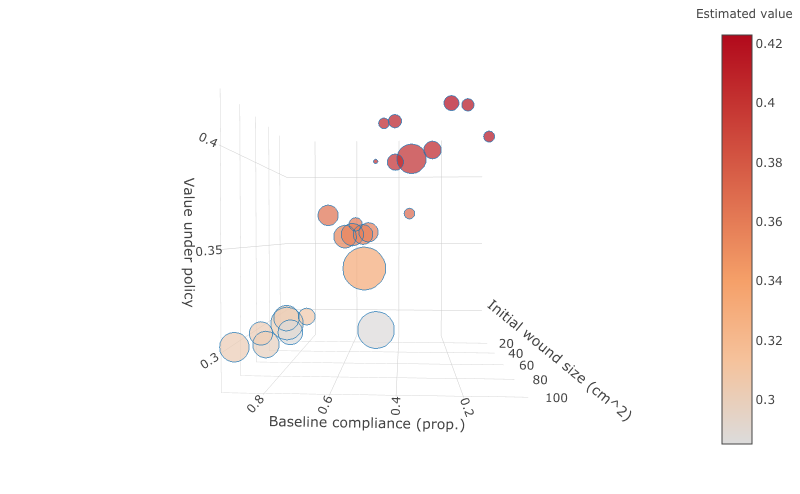}
    \label{fig:3dPlots}
    \caption{Two views of the values of policies evaluated during optimization}
\end{figure}

In addition to learning an optimal dynamic treatment regime, the goal of our analysis included characterizing the class of regimes under investigation to aid in translation of the information encoded in and proximal to optimal dynamic treatment regimes estimation to those making clinical decisions, clinicians with their patients. That is, we sought to exploit information learned through the optimal dynamic treatment regime learning process to supplement and enrich our understanding of the estimated optimal DTR, the policy class more generally, and the role and relevance of partial compliance in treatment decision-making. To do this, we first visualized the policies evaluated during the expectation improvement step to understand how different policies relate in value and across the parameters using R version 4.0.4 and plotly version 4.10.0 (Figure \ref{fig:3dPlots}). Each point represents a policy in the policy parameter space. The axes denoted as ``Baseline compliance (prop.)" and ``Initial wound size ($\text{cm}^2$)" correspond to the policy threshold parameters $\theta_1$ and $\theta_2$, and the vertical axis denotes the estimated value for the corresponding policy. The size of the points corresponds to the precision of the estimate, the inverse of the standard deviation of draws from the posterior distribution of the value function. From inspection, it can be seen that there are a number of explored policies with similarly optimal values (nearly the same height in the z-axis) and that optimality of the the rule is closely tied to the policy parameter value for baseline compliance. We scaled the points in the figure to be proportional to the precision of the value estimate of the policy to understand which policies we were more sure of their value and which policies we were less certain of their value.

\begin{figure}
    \begin{center}
    \includegraphics[width = \textwidth]{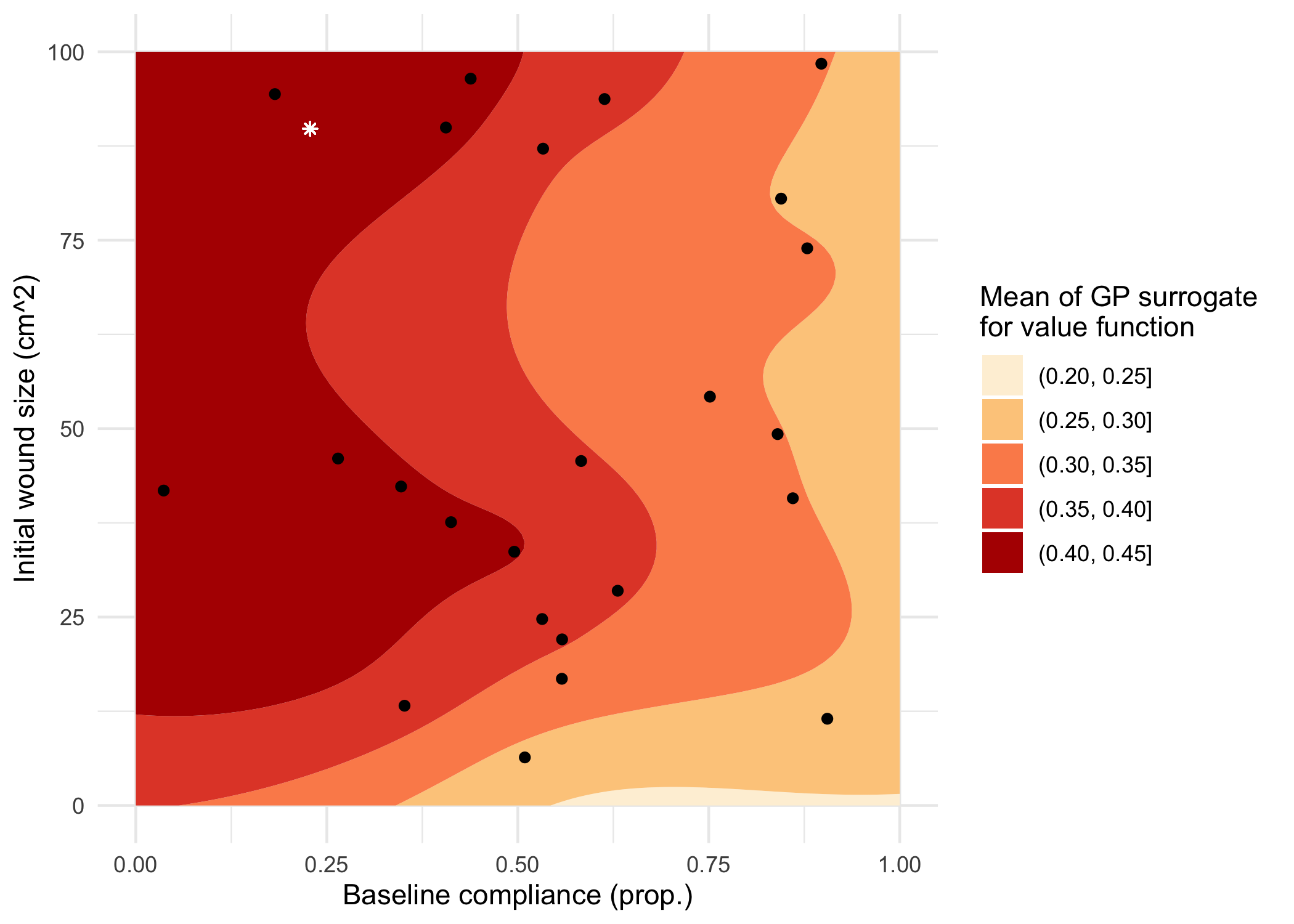}
    \label{fig:gpContours}
    \caption{Gaussian process surrogate mean for the value over the policy parameter space}
    \end{center}
\end{figure}

In addition analyzing the policies explored during the Bayesian optimization step, we also examined the mean of a Gaussian process surrogate for the value function over the parameter space. One could use directly the Gaussian process used for expectation improvement, although we, for ease of exposition, chose to fit a new Gaussian process with a Matern kernel with noise using the saved counterfactual draws from the Bayesian G-computation step. Across a $100 \times 100$ grid of the parameter space, we evaluated the mean of the fitted Gaussian process using Python version 3.9.7, scikit-learn version 0.24.2. We plotted the results (using R version 4.0.4, ggplot2 version 3.3.5), shown in Figure \ref{fig:gpContours}, as level sets represented by different colors where each level represents a 5\% range in the approximated proportion of individuals that would be healed, alive, and with their limb intact at 6 months if they were to follow the policy implied by the parameters for baseline compliance and initial wound size. We chose levels sets, as opposed to a continuous gradient, to identify policies with nearly equivalent clinically significant meaning. Each point represents a policy in the policy parameter space that was evaluated during the expectation improvement step. The estimated optimal policy is denoted by a white star. Strikingly, a swath of policies, those that recommend revascularization along with continued wound care when baseline compliance is less than about 35\% or initial wound size is greater than 12 cm$^2$, yield nearly equivalent approximated values suggesting that following a number of policies would lead to outcomes similar to that of the optimal policy. This means that clinicians and their patients may have some ``wiggle room'' in optimal decision making; it also shows under which policies near-optimality drops off. This latter point may be of particular importance for patients who may face external barriers (e.g. lack of transportation, child care challenges, work obligations, etc.) to completing their wound care appointments. Finally we note that  after hyperparameter tuning, the fitted Gaussian process has lengthscale 1.65 for the baseline compliance parameter and 103 for the initial wound size parameter ($\sigma^2 = 0.244$, $\nu = 1.5$, noise = 0.000274) providing some evidence that the class of policies of interest, the parameter for baseline compliance plays a bigger role in determining the value of the policy than the parameter for initial wound size. 

\section{Discussion}
\label{s:discuss}

In this paper, we describe a strategy for computationally efficient learning of optimal DTRs using Bayesian optimization methodology. Perhaps even more importantly, we attempt to narrow the gap between precision medicine statistical analyses and translatable clinical evidence by expanding the question set being asked and answered in precision medicine. Instead of focusing solely on finding optimal DTRs, we also show and argue for identifying policies that yield equally optimal results in terms of clinical significance and exploring how the value of policies change across policy parameters which may give insight into what features are driving the benefits gained from precision medicine decision support and which features are less important for prescribing treatments. By asking new questions and providing new types of evidence, we get us closer to realizing the promise of precision medicine, matching the right treatment to the right patient at the right time, in the real world. 

There are limitations of our work. For example, our approach hinges on having a good estimator for the value across the policy parameter space. Additionally, it is not clear how large the exploration budget should be for Bayesian optimization and guarantees of finding a good optimal policy depend on the underlying value function at least being sufficiently smooth. Finally, while simple policy classes are easy to understand and can be visualized well, picking the right form of policies and the size of the policy class remains a challenge.

A wide range of future work is motivated by the ideas we have proposed. For example, extending and analyzing the approach in the multi-stage setting or with more complex parametric decision rules such as trees and lists.  Moreover, our work is a first pass at expanding the targets of a precision medicine analysis to generate evidence to directly support translation to the clinic. Future work for the precision medicine field at large may include better understanding of how clinicians and patients use the type of evidence we have suggested generating and the questions and answers it inspires. 



\section*{Acknowledgements}
The first and fourth authors were supported  by a grant from the National Institute on Drug Abuse (NIDA) (No. R01 DA048764). The project described was supported by the National Center for Advancing Translational Sciences (NCATS), National Institutes of Health, through Grant Award Number UL1TR002489. The content is solely the responsibility of the authors and does not necessarily represent the official views of the NIH.
\vspace*{-8pt}


%
\bibliography{bibliography}

\appendix

\section{Identification of $\mathcal{V}(d)$}
In this section, we show how $\mathcal{V}(d)$ is identified from the observed data. Using a similar argument as in \citep{jiang2021}, observe that $\mathbb{E}[Y(a)\vert C_1(0)]$ for $a = 0, 1$ is identified from the observed data given that the SUTVA, NUC, PI, and OSPC assumptions hold and that the family of distributions $\{P(C_1(0)\vert X= x)\}$ is complete. In particular,
\begin{align*}
    \mathbb{E}[Y \vert A_1 = a, X = x] =& \mathbb{E}[Y(a) \vert X = x] & \text{(SUTVA, NUC)}\\
    =& \mathbb{E}[Y(a) \vert C_1(0),  X = x] & \text{(PI, OSPC)} \\
    =& \int_{0}^{1} \mathbb{E}[Y(a) \vert C_1(0) = c_1] dP(C_1(0) \vert X = x)
\end{align*}
is an integral equation that uniquely identifies $\mathbb{E}[Y(a)\vert C_1(0) = c_1]$ from the observed data. Identification of $\mathcal{V}(d)$ follows by noting that 
\begin{align*}
    \mathcal{V}(d(x)) =& \mathbb{E}\{\mathbb{E}[\mathbbm{1}(d(x) = 0)Y(0) + \mathbbm{1}(d(x) = 1) Y(1) \vert X = x]\}.
\end{align*}

\section{Additional simulation results}
\begin{table}
    \footnotesize
    \centering
    \caption{Mean squared error between value of optimal DTR and estimated optimal DTR.}
    \begin{tabular}{c|c|l|l|l|l|l|l|l|l}
         & & \multicolumn{2}{|c|}{IPW} & \multicolumn{2}{|c|}{sIPW} & \multicolumn{2}{|c|}{G-comp} &\multicolumn{2}{|c|}{AIPWE} \\
         N & CV or $w$ & MSE & MC error & MSE & MC error & MSE & MC error & MSE & MC error\\ \hline 
        \multicolumn{10}{c}{Simulation Setting 1} \\ \hline
        \multirow{3}{*}{200} & 0.75 & 0.0043 & 0.0002 & 0.0005 & 0.0000 & 0.0002 & 0.0000 & 0.0002 & 0.0000 \\ 
         & 1.0 & 0.0057 & 0.0002 & 0.0008 & 0.0000 & 0.0004 & 0.0000 & 0.0004 & 0.0000 \\ 
         & 1.25 & 0.0073 & 0.0003 & 0.0014 & 0.0001 & 0.0007 & 0.0000 & 0.0007 & 0.0000 \\ 
        \multirow{3}{*}{500} & 0.75 & 0.0017 & 0.0001 & 0.0002 & 0.0000 & 0.0001 & 0.0000 & 0.0001 & 0.0000 \\ 
         & 1.00 & 0.0022 & 0.0001 & 0.0004 & 0.0000 & 0.0002 & 0.0000 & 0.0002 & 0.0000 \\ 
         & 1.25 & 0.0032 & 0.0001 & 0.0006 & 0.0000 & 0.0003 & 0.0000 & 0.0003 & 0.0000 \\ 
        \multirow{3}{*}{1000} & 0.75 & 0.0009 & 0.0000 & 0.0001 & 0.0000 & 0.0000 & 0.0000 & 0.0000 & 0.0000 \\ 
         & 1.00 & 0.0011 & 0.0000 & 0.0002 & 0.0000 & 0.0001 & 0.0000 & 0.0001 & 0.0000 \\ 
         & 1.25 & 0.0015 & 0.0001 & 0.0003 & 0.0000 & 0.0001 & 0.0000 & 0.0001 & 0.0000 \\ \hline
        \multicolumn{10}{c}{Simulation Setting 2} \\ \hline 
        \multirow{3}{*}{200}& 0.75 & 0.0008 & 0.0000 & 0.0007 & 0.0000 & 0.0048 & 0.0002 & 0.0050 & 0.0002 \\ 
         & 1.00 & 0.0029 & 0.0001 & 0.0028 & 0.0001 & 0.0722 & 0.0009 & 0.0029 & 0.0001 \\ 
         & 1.25 & 0.0052 & 0.0002 & 0.0038 & 0.0002 & 0.0833 & 0.0009 & 0.0061 & 0.0002 \\ 
        \multirow{3}{*}{500} & 0.75 & 0.0004 & 0.0000 & 0.0004 & 0.0000 & 0.0024 & 0.0001 & 0.0025 & 0.0001 \\ 
         & 1.00 & 0.0014 & 0.0001 & 0.0014 & 0.0001 & 0.0793 & 0.0006 & 0.0013 & 0.0001 \\ 
         & 1.25 & 0.0029 & 0.0001 & 0.0016 & 0.0001 & 0.0835 & 0.0006 & 0.0042 & 0.0001 \\ 
        \multirow{3}{*}{1000} & 0.75 & 0.0018 & 0.0002 & 0.0004 & 0.0000 & 0.0016 & 0.0001 & 0.0016 & 0.0001 \\ 
         & 1.00 & 0.0010 & 0.0001 & 0.0009 & 0.0001 & 0.0846 & 0.0005 & 0.0009 & 0.0001 \\ 
         & 1.25 & 0.0014 & 0.0001 & 0.0010 & 0.0000 & 0.0855 & 0.0004 & 0.0040 & 0.0001 \\ \hline
        \multicolumn{10}{c}{Simulation Setting 3} \\ \hline
        \multirow{3}{*}{200} & 0.75 & 0.0045 & 0.0002 & 0.0047 & 0.0002 & 0.0254 & 0.0006 & 0.0047 & 0.0002 \\ 
         & 1.00 & 0.0031 & 0.0002 & 0.0032 & 0.0002 & 0.0151 & 0.0004 & 0.0032 & 0.0002 \\ 
         & 1.25 & 0.0040 & 0.0002 & 0.0042 & 0.0002 & 0.0118 & 0.0003 & 0.0045 & 0.0002 \\ 
        \multirow{3}{*}{500}& 0.75 & 0.0039 & 0.0001 & 0.0040 & 0.0001 & 0.0254 & 0.0005 & 0.0027 & 0.0001 \\ 
         & 1.00 & 0.0013 & 0.0001 & 0.0013 & 0.0001 & 0.0176 & 0.0003 & 0.0013 & 0.0001 \\ 
         & 1.25 & 0.0015 & 0.0001 & 0.0015 & 0.0001 & 0.0105 & 0.0003 & 0.0019 & 0.0001 \\ 
        \multirow{3}{*}{1000} & 0.75 & 0.0038 & 0.0001 & 0.0039 & 0.0001 & 0.0246 & 0.0004 & 0.0020 & 0.0001 \\ 
         & 1.00 & 0.0008 & 0.0000 & 0.0007 & 0.0000 & 0.0193 & 0.0002 & 0.0007 & 0.0000 \\ 
         & 1.25 & 0.0009 & 0.0000 & 0.0008 & 0.0000 & 0.0112 & 0.0002 & 0.0010 & 0.0001 \\ 
    \end{tabular}
    \label{tab:sim1}
\end{table}

\begin{table}
    \footnotesize
    \centering
    \caption{Difference between the true value function and the surrogate value function.}
    \begin{tabular}{ccllllllll}
         &  & \multicolumn{2}{c}{IPW} & \multicolumn{2}{c}{sIPW} & \multicolumn{2}{c}{G-computation} & \multicolumn{2}{c}{AIPWE}\\
        N & $\sigma$ & $\Vert \cdot \Vert_{1}$ & $\Vert \cdot \Vert_{2}$ & $\Vert \cdot \Vert_{1}$ & $\Vert \cdot \Vert_{2}$  & $\Vert \cdot \Vert_{1}$ & $\Vert \cdot \Vert_{2}$ & $\Vert \cdot \Vert_{1}$ & $\Vert \cdot \Vert_{2}$\\ \hline
        \multicolumn{10}{c}{Simulation Setting 1} \\ \hline
        \multirow{3}{*}{200} & 0.75 & 0.0499 & 0.0569 & 0.0272 & 0.0351 & 0.0210 & 0.0278 & 0.0210 & 0.0277 \\ 
         & 1.00 & 0.0594 & 0.0660 & 0.0326 & 0.0404 & 0.0243 & 0.0303 & 0.0243 & 0.0303 \\ 
         & 1.25 & 0.0693 & 0.0752 & 0.0391 & 0.0466 & 0.0285 & 0.0338 & 0.0285 & 0.0338 \\ 
        \multirow{3}{*}{500} & 0.75 & 0.0332 & 0.0390 & 0.0196 & 0.0262 & 0.0161 & 0.0224 & 0.0161 & 0.0224 \\ 
         & 1.00 & 0.0376 & 0.0427 & 0.0224 & 0.0283 & 0.0173 & 0.0222 & 0.0173 & 0.0222 \\ 
         & 1.25 & 0.0459 & 0.0501 & 0.0262 & 0.0315 & 0.0196 & 0.0236 & 0.0196 & 0.0236 \\ 
        \multirow{3}{*}{1000} & 0.75 & 0.0252 & 0.0306 & 0.0160 & 0.0223 & 0.0138 & 0.0201 & 0.0138 & 0.0201 \\ 
         & 1.00 & 0.0280 & 0.0323 & 0.0171 & 0.0221 & 0.0138 & 0.0185 & 0.0138 & 0.0185 \\ 
         & 1.25 & 0.0323 & 0.0357 & 0.0191 & 0.0235 & 0.0150 & 0.0187 & 0.0150 & 0.0187 \\ \hline
        \multicolumn{10}{c}{Simulation Setting 2} \\ \hline
        \multirow{3}{*}{200}& 0.75 & 0.0431 & 0.0492 & 0.0402 & 0.0461 & 0.3759 & 0.3983 & 0.3631 & 0.3878 \\ 
         & 1.00 & 0.0567 & 0.0624 & 0.0568 & 0.0628 & 0.1051 & 0.1311 & 0.0654 & 0.0724 \\ 
         & 1.25 & 0.0698 & 0.0742 & 0.0656 & 0.0716 & 0.1713 & 0.1856 & 0.1296 & 0.1385 \\ 
        \multirow{3}{*}{500}& 0.75 & 0.0282 & 0.0339 & 0.0273 & 0.0329 & 0.3758 & 0.3977 & 0.3626 & 0.3867 \\ 
         & 1.00 & 0.0382 & 0.0432 & 0.0384 & 0.0436 & 0.0889 & 0.1185 & 0.0445 & 0.0502 \\ 
         & 1.25 & 0.0455 & 0.0521 & 0.0416 & 0.0463 & 0.1647 & 0.1777 & 0.1182 & 0.1264 \\ 
        \multirow{3}{*}{1000} & 0.75 & 0.0311 & 0.0426 & 0.0208 & 0.0266 & 0.3766 & 0.3983 & 0.3612 & 0.3852 \\ 
         & 1.00 & 0.0284 & 0.0333 & 0.0284 & 0.0336 & 0.0813 & 0.1144 & 0.0319 & 0.0371 \\ 
         & 1.25 & 0.0317 & 0.0354 & 0.0302 & 0.0344 & 0.1653 & 0.1772 & 0.1188 & 0.1270 \\ \hline
        \multicolumn{10}{c}{Simulation Setting 3} \\ \hline 
        \multirow{3}{*}{200}& 0.75 & 0.0578 & 0.0659 & 0.0576 & 0.0660 & 0.0903 & 0.1041 & 0.0785 & 0.0879 \\ 
         & 1.00 & 0.0590 & 0.0661 & 0.0575 & 0.0646 & 0.0779 & 0.0904 & 0.0677 & 0.0752 \\ 
         & 1.25 & 0.0619 & 0.0679 & 0.0639 & 0.0698 & 0.0809 & 0.0909 & 0.0737 & 0.0798 \\ 
        \multirow{3}{*}{500} & 0.75 & 0.0401 & 0.0483 & 0.0404 & 0.0486 & 0.0766 & 0.0902 & 0.0611 & 0.0694 \\ 
         & 1.00 & 0.0406 & 0.0471 & 0.0403 & 0.0470 & 0.0597 & 0.0726 & 0.0452 & 0.0518 \\ 
         & 1.25 & 0.0414 & 0.0467 & 0.0417 & 0.0470 & 0.0608 & 0.0708 & 0.0515 & 0.0568 \\ 
        \multirow{3}{*}{1000} & 0.75 & 0.0321 & 0.0406 & 0.0320 & 0.0405 & 0.0714 & 0.0848 & 0.0548 & 0.0625 \\ 
         & 1.00 & 0.0308 & 0.0376 & 0.0302 & 0.0371 & 0.0500 & 0.0645 & 0.0339 & 0.0404 \\ 
         & 1.25 & 0.0310 & 0.0363 & 0.0310 & 0.0364 & 0.0500 & 0.0601 & 0.0398 & 0.0449 \\ 
    \end{tabular}
    \label{tab:sim2}
\end{table}

\end{document}